\documentclass[journal=jacsat,manuscript=article]{achemso}
\usepackage[version=3]{mhchem}
\usepackage{hyperref}
\usepackage{xcolor}

\author{Laura Borgongino}
\affiliation{NEST, Istituto Nanoscienze-CNR and Scuola Normale Superiore, I-56127, Pisa, Italy}
\email{laura.borgongino@sns.it}
\author{Rubén Seoane Souto}
\affiliation{Instituto de Ciencia de Materiales de Madrid (ICMM), Consejo Superior de Investigaciones Científicas (CSIC), Sor Juana Inés de la Cruz 3, 28049 Madrid, Spain}
\author{Alessandro Paghi}
\affiliation{NEST, Istituto Nanoscienze-CNR and Scuola Normale Superiore, I-56127, Pisa, Italy}
\author{Giulio Senesi}
\affiliation{NEST, Istituto Nanoscienze-CNR and Scuola Normale Superiore, I-56127, Pisa, Italy}
\author{Katarzyna Skibinska}
\affiliation{NEST, Istituto Nanoscienze-CNR and Scuola Normale Superiore, I-56127, Pisa, Italy}
\author{Lucia Sorba}
\affiliation{NEST, Istituto Nanoscienze-CNR and Scuola Normale Superiore, I-56127, Pisa, Italy}
\author{Elisa Riccardi}
\affiliation{NEST, Istituto Nanoscienze-CNR and Scuola Normale Superiore, I-56127, Pisa, Italy}
\author{Francesco Giazotto}
\affiliation{NEST, Istituto Nanoscienze-CNR and Scuola Normale Superiore, I-56127, Pisa, Italy}
\author{Elia Strambini}
\affiliation{NEST, Istituto Nanoscienze-CNR and Scuola Normale Superiore, I-56127, Pisa, Italy}
\email{elia.strambini@cnr.it}

\title[Biharmonic-Drive Tunable Josephson Diode]
  {Biharmonic-drive tunable Josephson diode}

\begin{document}

\begin{abstract}
The superconducting diode effect has garnered significant interest due to its prospective applications in cryogenic electronics and computing, enabling directional supercurrent transport. This phenomenon has been demonstrated across various superconducting platforms, which typically necessitate unconventional materials with broken spatial symmetries or external magnetic fields, posing scalability and integration challenges. This work introduces an innovative method to realize the superconducting diode effect by disrupting spatio-temporal symmetries in a conventional Josephson junction utilizing a biharmonic AC drive signal. We achieve wireless modulation of the diode's polarity and efficiency with an antenna. Our findings indicate a diode efficiency reaching the ideal 100\% over a broad frequency range, with a temperature resilience up to 800 mK, and efficient AC signal rectification. This versatile and platform-independent superconducting diode signifies a promising component for advancing future superconducting digital electronics, including efficient logic gates, ultra-fast switches, and dynamic half-wave supercurrent rectifiers.
\end{abstract}
\vspace{1em}
\noindent\textbf{Keywords:} Josephson Junctions, Superconducting diode effect, Josephson diode, Biharmonic drive, Superconducting electronics.

\vspace{1em}
\noindent
The superconducting diode effect has been extensively studied in recent years \cite{nadeem_superconducting_2023,moll_evolution_2023} due to its potential applications in cryogenic electronics and dissipationless computation. Its zero resistance state, and thus zero energy loss, makes the superconducting diode a promising building block for future superconducting technologies\cite{holmes_energy-efficient_2013}, for ultra-low power consumption electronics and high-frequency rectifiers \cite{borodianskyi_josephson_2017}. In fact, due to their nonreciprocal current-voltage characteristic, superconducting diodes can work as AC signal rectifiers, the core of signal processing and power supply conversion.
They have been implemented in various platforms that break spatial symmetry either via non-centrosymmetric materials \cite{ando_observation_2020,narita_field-free_2022,narita_superconducting_2024} or engineered in heterostructures \cite{ideue_symmetry_2021,baumgartner_supercurrent_2022,cai_superconductorferromagnet_2023}. The superconducting diode effect was demonstrated in different spin-orbit interaction-based platforms under the action of a magnetic field \cite{ando_observation_2020,turini_josephson_2022,baumgartner_supercurrent_2022, bauriedl_supercurrent_2022}, in field-free Josephson junctions (JJs)\cite{wu_field-free_2022,narita_field-free_2022,yun_magnetic_2023,strambini_superconducting_2022}, graphene-based systems\cite{diez-merida_symmetry-broken_2023,lin_zero-field_2022}, SQUIDs, \cite{paolucci_gate-_2023,valentini_parity-conserving_2024} and gate-tunable structures \cite{gupta_gate-tunable_2023,ciaccia_gate-tunable_2023,mazur_gate-tunable_2024}.
Superconducting diodes were also made by integrating a superconductor with a magnetic atom on top of it \cite{trahms_diode_2023}, trapping the Abrikosov vortex \cite{golod_demonstration_2022}, and Meissner screening \cite{hou_ubiquitous_2023}. Lately, high-temperature superconducting diodes have been proposed, based on JJs made of twisted van der Waals bilayer \cite{ghosh_high-temperature_2024} or by cuprates \cite{qi_high-temperature_2025}.
However, most of these devices require specially designed materials and heterostructures with broken spatial symmetries or the application of a magnetic field to break time reversal, making them difficult to scale and demanding to integrate into computer chips and quantum devices. Moreover, one of the significant limitations of superconducting diodes is their typically low rectification efficiency. To overcome this limitation, various strategies have been explored, including three-terminal JJ \cite{chiles_nonreciprocal_2023}, radio-frequency-driven-Josephson diodes \cite{seoane_souto_tuning_2024,valentini_parity-conserving_2024, cuozzo_microwave-tunable_2024}, asymmetric interferometric devices \cite{souto_josephson_2022,Bozkurt_SciPost2023}, and periodically driven quantum-dot-based systems \cite{soori_nonequilibrium_2023,ortega-taberner_anomalous_2023}. Here, we demonstrate a superconducting diode based on broken spatio-temporal symmetries, induced by an external multi-harmonic AC drive\cite{scheer_tunable_2025}, applied to a conventional JJ. We experimentally implement such a scheme in an InAs-based JJ driven by a biharmonic drive: an AC drive signal composed of two frequency components, which can be galvanically injected into the JJ or irradiated by an antenna. 
The phase shift between the two harmonics controls the AC drive's asymmetry and tunes the superconducting diode direction and efficiency. The antenna enables wireless control of the device, making the diode scalable, easy to reconfigure, and suitable for logic operations due to the fast tunability\cite{hosur_digital_2024}. Furthermore, since this scheme is independent of the specific design or structure of the device, our superconducting diode is compatible with various JJs platforms.

\begin{figure}
    \centering
    \includegraphics[width=0.85\linewidth]{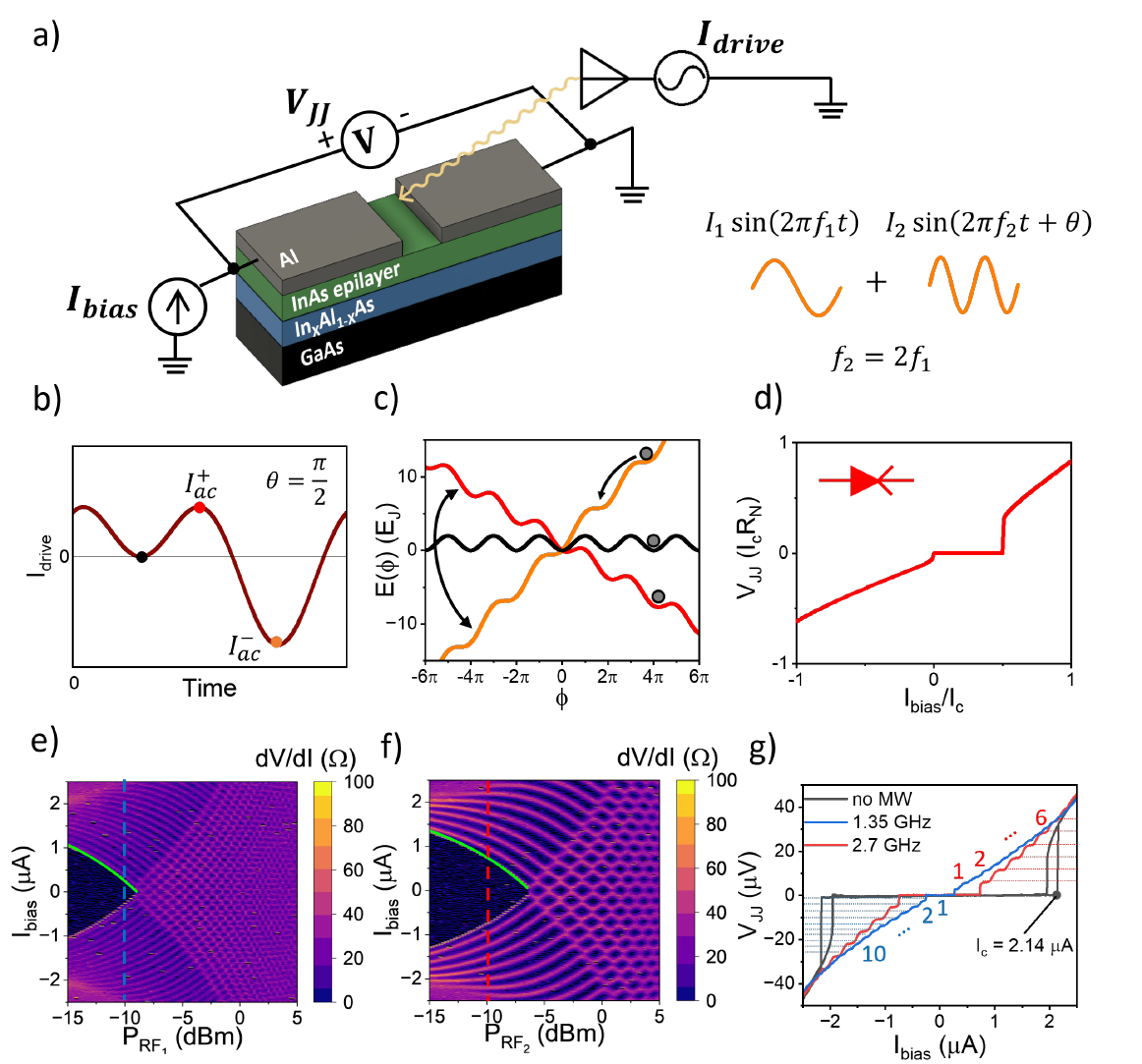}
    \caption{\textbf{Concept device and pre-characterization.} a) Schematic structure of the Al-InAs JJ and conceptual sketch of the setup. b) Biharmonic drive signal applied to the antenna at a phase shift \(\theta =\pi /2\). c) Dynamics of the washboard potential. Different colors refer to the selected time in b). The tilt of the washboard potential evolves according to the instantaneous amplitude of the AC drive: zero (black curve), maximum $I_{ac}^+$ (red curve), and minimum  $I_{ac}^-$ (orange curve). The gray dot represents the phase particle. d) VI characteristics simulated from the drive in (b): at positive bias, below the critical current, the device is superconducting, while at negative bias, there is a non-zero voltage drop across the JJ. We set t $R_N=10$ $\hbar/2e^2$, $C=0.02$ $eI_c/\hbar$, $2\pi f_1=0.05\omega_c$ and $I_1=2I_2$.  e)-f) Differential resistance measured as a function of the current bias (\(I_{bias}\)) and applied power ($P_{RF}$) at \(f=1.35\) GHz and \(f=2.7\) GHz showing the characteristic Shapiro plateaus. Green curves represents the fits of $I_c(P_{RF})=I_c(0)-Z(\omega)\sqrt{P_{RF}}$ with \(I_c(0)=2.14\) \(\mu\)A, $Z(1.35 \text{ GHz})=6$ $\mu\text{A}/\sqrt{\text{mW}}$ and $Z(2.7 \text{ GHz})=4.4 \, \mu\text{A}/\sqrt{\text{mW}}$. g) VI curves with drive at \(P_{RF}=-10\) dBm and \(f=1.35 \) GHz and \(2.7\) GHz (blue and red, respectively corresponding to the dashed lines in (f)) and no microwave applied (gray). The numbers denote the Shapiro steps observed at $\Delta V = n f \frac{h}{2e}$. From the gray curve, we estimate \(I_c=2.14\) $\mu$A, \(R_N\approx26\) $\Omega$ and \(I_cR_N\approx55\) $\mu$V. Measurements are taken at \(T\approx70\) mK.}
    \label{fig:Figure1}
\end{figure}
A sketch of the device, together with the diode idea and a first electrical characterization, is presented in Figure \ref{fig:Figure1}.
As shown in Figure \ref{fig:Figure1}a, the device is placed in a four-terminal configuration under a microwave irradiation (AC drive) provided by a broadband antenna placed a few millimeters away from the chip surface, as conventionally used in Shapiro experiments \cite{larson_zero_2020,arnault_multiterminal_2021,iorio_half-integer_2023}.
Unlike the latter, we excite the antenna with a biharmonic signal composed by a superposition of two tones with different frequencies, amplitudes, and phases. The radiation emitted by the antenna is absorbed by the JJ, inducing an AC current of the form:
\begin{equation}
I_{drive}(t)=I_1\sin{(2\pi f_1t)}+I_2\sin{(2\pi f_2t+\theta)},
\label{Eq1}
\end{equation}
where \(f_2=2f_1\), and  \(I_1,I_2>0\) are the current amplitudes induced by the two tones.  The phase shift \(\theta\) can be adjusted so that the maximum \((I_{ac}^+)\) and the minimum \((I_{ac}^-)\) of the AC drive are different (asymmetric drive) $I_{ac}^+\neq |I_{ac}^-|$, that is a sufficient condition to obtain non-reciprocity in the JJ\cite{scheer_tunable_2025} with a maximum asymmetry obtained for \(\theta=\pi /2+n\pi\) (Figure \ref{fig:Figure1}b).
The VI characteristics of the JJ can be described by the phase dynamics with the resistively and capacitively shunted junction (RCSJ) model\cite{stewart_currentvoltage_1968,mccumber_effect_1968}:
\begin{equation}
\frac{C\hbar}{2e}\ddot{\phi}+ \frac{\hbar}{2eR_N}\dot{\phi}+I(\phi)=I_{drive}(t)+I_{bias},
\label{Eq2}
\end{equation}
where $\phi$ is the phase, \(C\) and $R_N$ are the capacitance and normal-state resistance of the JJ, $\hbar$ is the reduced Planck constant and $e$ is the electronic charge. \(I(\phi)=I_c\sin(\phi)\) is the current-phase relation (CPR), with \(I_c\) being the critical current of the JJ, and \(I_{bias}\) is a DC current bias.
Within this model, the phase is a classical variable, equivalent to a particle under the action of a washboard potential that tilts in time according to the external drive: when \(|I_{drive}|<I_c\) the phase particle is confined in a local minimum whereas when \(|I_{drive}|> I_c\), it is free to roll down the washboard potential resulting in a non-zero voltage drop \(V=\hbar \dot{\phi}/2e\). 
If the drive is asymmetric in direction, it is possible to achieve a non-zero voltage drop during the AC cycle only in one direction, if \(I_{ac}^+< I_c<|I_{ac}^-|\), as shown in Figure \ref{fig:Figure1}c. As a result, the VI curve of the JJ becomes asymmetric up to an ideal supercurrent diode effect as illustrated in Figure \ref{fig:Figure1}d, where supercurrent can flow only in one direction. Within the adiabatic approximation \(2\pi f_1 \ll \omega_c\), where \(\omega_c=2eI_cR_N/\hbar\), it is easy to show that the maximum supercurrent allowed in the forward/backward direction ($I_c^+$ and $I_c^-$, respectively) follow the simple relation \(I_c^\pm=\pm(I_c-|I_{ac}^\pm|)\)~\cite{tinkham_introduction_2004}. When the diode effect occurs, the critical currents with positive or negative bias are different, namely \(I_c^+\ne |I_c^-|\).
Obtaining ideal diode requires \(\text{max}(I_{ac}^+,|I_{ac}^-|)=I_c\). For the driving current in Eq. (\ref{Eq1}), this condition is achieved for \(|I_1|+|I_2|=I_c\) and \(\theta=\pi /2+n\pi\). To quantify the diode effect, we use the diode efficiency \(\eta\) defined as \(\eta =\frac{I_c^+-|I_c^-|}{I_c^++|I_c^-|}\) that equals to 0 when there is no diode effect and $\pm1$ in the ideal case (forward and backward diode, respectively), corresponding to $I_{c}^-=0$ or $I_{c}^+=0$. Within the adiabatic approximation, we obtain \(\eta =\frac{I_{ac}^+-|I_{ac}^-|}{I_{ac}^++|I_{ac}^-|-2I_c}\), that can be simplified into:
\begin{equation}
\eta=\frac{\eta_{ac}}{1-\frac{2I_c}{I_{ac}^++|I_{ac}^-|}},
\label{Eq3}
\end{equation}
where \(\eta_{ac} =\frac{I_{ac}^+-|I_{ac}^-|}{I_{ac}^++|I_{ac}^-|}\) quantifies the asymmetry of the drive. The equation shows that even for non-fully asymmetric drives ($\eta_{ac}<1$), it is possible to achieve ideal diodes ($\eta=1$). Moreover, the latter expression is valid for any adiabatic drive beyond the biharmonic one and generic non-sinusoidal JJs. 
In the following, we implement this scheme for a semiconducting Al-InAs JJ fabricated on the InAs on Insulator (InAsOI) platform \cite{paghi_inas_2025,paghi_josephson_2025,paghi_supercurrent_2024,battisti_extremely_2024}.
The device is made of a 350 \(\mu\)m-thick GaAs (100) substrate, a 50 nm GaAs buffer, a 100 nm GaAs/AlGaAs superlattice, a 50 nm GaAs layer, a 1.25 \(\mu\)m-thick \(In_{X}Al_{1-X}As\) metamorphic buffer layer (X from 0.15 to 0.81), a 100 nm-thick InAs intrinsically n-doped epilayer and two Al leads (100 nm of thickness and 5.6 \(\mu\)m of width) with interelectrode separation of 480 nm. Fabrication details can be found in previous works based on the same platform \cite{paghi_inas_2025,paghi_josephson_2025,paghi_supercurrent_2024,battisti_extremely_2024}.
To characterize the device and to ensure a good coupling between the microwave radiation and the JJ, a single drive with frequency in the range of GHz is first applied to the antenna, showing clear Shapiro steps at different injected RF powers ($P_{RF}$) and frequencies ($f$).
Technical information about the experimental setup is reported in the Supporting Information, Methods section. Figures \ref{fig:Figure1}e and \ref{fig:Figure1}f, show the differential resistance $dV/dI$ vs the DC bias current ($I_{bias}$) and $P_{RF}$ measured at 1.35 GHz and 2.7 GHz frequencies, respectively. In the two cases, the characteristic Shapiro steps are visible in the zeros of the differential resistance represented by the darker regions of the plots. The corresponding plateaus are visible in Figure \ref{fig:Figure1}g, showing the VI characteristics measured at \(P_{RF}=-10\) dBm for the two frequencies and compared to the gray curve obtained without microwave radiation. It is possible to estimate \(\omega_c = 2\pi\times26\) GHz from the latter, confirming that the drive is in the adiabatic regime. Precise integer Shapiro steps appear at voltage \(n f \frac{h}{2e}\) with \(n\) ranging from -10 to 10 (1.35 GHz) and from -6 to 6 (2.7 GHz); half-integer Shapiro steps \cite{guarcello_probing_2024} are not observed, unveiling a negligible contribution of high order harmonics in the CPR \cite{iorio_half-integer_2023,valentini_parity-conserving_2024}. Shapiro steps were visible up to 900 mK, and at different frequencies ranging from 0.5 GHz to 3 GHz, as reported in the Supporting Information section Methods and Additional Data. The damping of $I_c(P_{RF})$ is a consequence of the induced AC current and follows a square root dependence in the power $I_c(P_{RF})=I_c(0)-I_{ac}(P_{RF})=I_c(0)-Z\sqrt{P_{RF}}$ as highlighted by the green fits superimposed to the two colorplots. It is possible to extract the coupling parameter $Z(f)$ from the fits, which enables the conversion between the injected microwave power $P_{RF}$ and the net AC current in the JJ.
\begin{figure}
    \centering
    \includegraphics[width=1\linewidth]{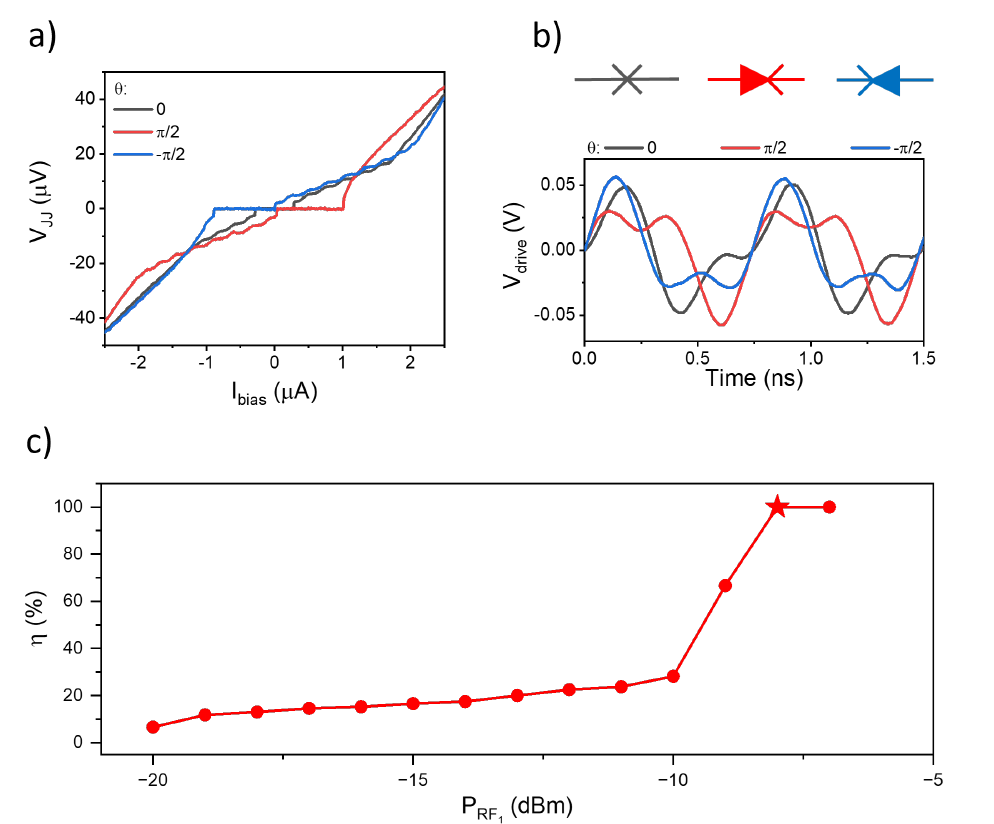}
    \caption{\textbf{Biharmonic drive diode.} a) VI curve under applied biharmonic drive (Eq.\ref{Eq1}) with \(f_1=1.35\) GHz and \(f_2=2.7\) GHz at different \(\theta \), \(P_{RF_1}=-8\) dBm and \(P_{RF_2}=-10\) dBm. b) Diode symbols and drive voltage monitored from the generator output: no diode (gray curve), positive diode (red curve), and negative diode (blue curve). c) Diode efficiency at different \(P_{RF_1}\), \(P_{RF_2}=-10\) dBm and $\theta=\pi/2$. The star corresponds to the $\eta$ estimated from the red curve in (a). Measurements are taken at \(T\approx70\) mK.}
    \label{fig:Figure2}
\end{figure}

The reciprocal VIs shown in Figure \ref{fig:Figure1}g evolve into a non-reciprocal one under the simultaneous action of the two microwave signals. This is demonstrated in Figure \ref{fig:Figure2}a, which illustrates the VI curves measured under the action of the biharmonic microwave signal (Eq. \ref{Eq1}) with $\theta=\pm \pi/2$. As predicted by the model, \(\theta\) determines the direction and efficiency of the diode: it can be controlled from ideal ($\eta=\pm1$ in $\theta=\pm \pi/2$) to not diode ($\eta=0$ in $\theta=0$). The ideal diode was obtained by adjusting \(\theta\) and the amplitude of the microwave drive, in good agreement with what was predicted in Ref \cite{scheer_tunable_2025}. Moreover, signatures of the two combined Shapiro steps are still visible at finite voltages. It is worth noting that, due to the asymmetry of the drive, Shapiro steps are mainly visible on the resistive side of the diode, where the microwave signal is stronger. This asymmetric behavior is further illustrated in Figure 4 of the Supporting Information, showing Shapiro steps at different combination of $P_{RF_1},P_{RF_2}$.
The driving signal was also monitored from the source at different $\theta$, confirming the biharmonic shape of the drive, as shown in Figure \ref{fig:Figure2}b.
In Figure \ref{fig:Figure2}c we show the evolution of $\eta$ at different \(P_{RF_1}\), while keeping $P_{RF_2}$ fixed and $\theta=\pi/2$. Notably, the ideal diode behavior occurs at $P_{RF_1}=-7$ dBm, corresponding to the condition $|I_{ac}^-(P_{RF_1},P_{RF_2})|=I_c$.
$\eta$ decreases monotonically by lowering $P_{RF_1}$, while maintaining the condition $|I_{ac}^-(P_{RF_1},P_{RF_2})|<I_c$. For values above $-7$ dBm, the drive amplitude becomes so large that the supercurrent was suppressed. The diode effect was tested at different combinations of \(P_{RF_1}\) and \(P_{RF_2}\) and the behavior of the diode efficiency as a function of the driving amplitudes was computed, as reported in Shapiro steps with a biharmonic drive and Diode efficiency at different $I_{1,2}$ subsections of the Supporting Information.
\begin{figure}
    \centering
    \includegraphics[width=1\linewidth]{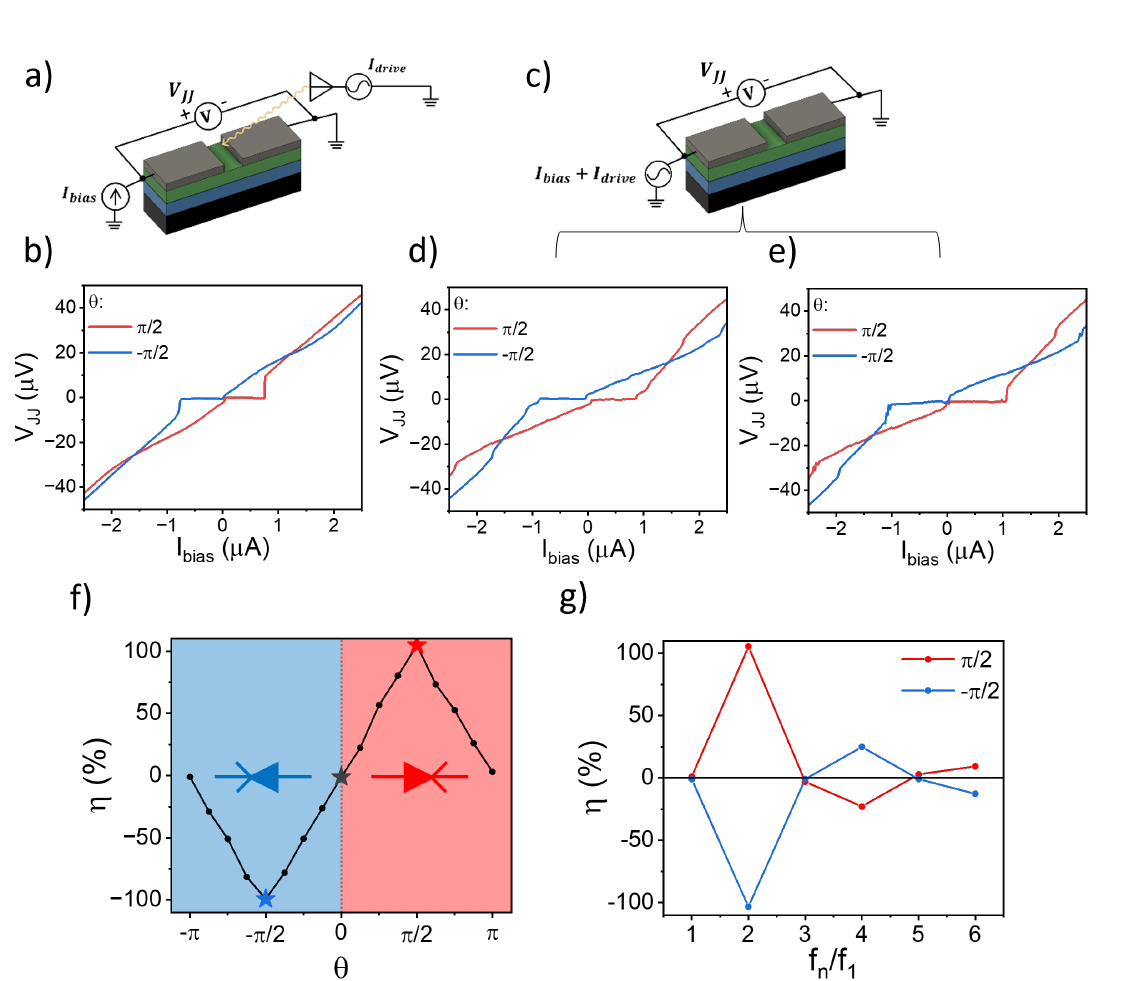}
    \caption{\textbf{Diode with biharmonic drive at different frequencies.} a) Sketch of the experimental setup with the driving signal injected through the antenna, with $P_{RF_1}=P_{RF_2}=5.75$ dBm. b) VI curve under an applied biharmonic drive with \(f_1=700\)  MHz at \(\theta=\pm\pi /2\). c) Sketch of the experimental setup with the driving signal injected through the JJ. $I_1=I_2$ and different peak-to-peak amplitude $I^{pp}\equiv I_{ac}^+-I_{ac}^-$. d-e) VI curve under applied biharmonic drive with \(\theta=\pm\pi /2\) , \(f_1=1\) kHz ($I^{pp}=4.4$ $\mu$A) and \(f_1=100\) Hz ($I^{pp}=3.8$ $\mu$A), respectively. f) Diode efficiency vs \(\theta\) for a low-frequency biharmonic drive used in (f). The blue and red parts refer to the direction of the diode. Colored stars indicate the ideal cases in panel (f). g) Diode efficiency at different \(f_n/f_1\) ($f_1=100$ Hz drive and $I^{pp}=3.8$ $\mu$A) for \(\theta=\pi /2\)  (red curve) and  \(\theta=-\pi /2\) (blue curve). Measurements are taken at \(T\approx70\) mK.}
    \label{fig:Figure3}
\end{figure}
\newline
Observing the diode effect with a biharmonic drive is also possible on a different platform, such as a SNS JJ, as reported in Supporting Information, where non-reciprocal VI curves for a Nb/Au/Nb JJ are presented.
Additionally, the biharmonic-driven diode effect is not limited to the GHz frequency range, where Shapiro steps can be observed, but is visible within a broader range of frequencies.
Indeed, we test it at different drive frequencies within the adiabatic approximation, ranging from 100 Hz to a few GHz.
In Figure \ref{fig:Figure3}a, a MHz drive is applied through the antenna: we observe a tunable diode effect at \(f_1=700\) MHz with negligible Shapiro steps, as shown in Figure \ref{fig:Figure3}b. To prove the validity of the effect also at lower frequencies (\(f_1\ll\)MHz), the driving signal is galvanically injected into the junction according to the scheme presented in Figure \ref{fig:Figure3}c, since the physical dimensions of the antenna limit the frequency bandwidth to 0.5-5 GHz (see Supporting Information, Methods section).
Adjusting the phase shift and the amplitude of the signal, we also achieve $\eta=1$ for \(f_1=1\) kHz and \(f_1=100\) Hz as shown in Figure \ref{fig:Figure3}d and \ref{fig:Figure3}e, respectively.
In Figure \ref{fig:Figure3}f, $\eta$ is quantified as a function of \(\theta\) for the 100 Hz biharmonic driving signal, demonstrating the continuous tunability of the effect.
Moreover, the effect of the diode is not limited by the choice of the second harmonic ($f_2=2f_1$). Still, it can be easily demonstrated that asymmetric drives are possible for all even harmonics ($f_{2n}=2n f_1$). In Figure \ref{fig:Figure3}g we evaluate $\eta$ obtained at different $f_{2n}$ for fixed $I_1, I_2$ optimized for  \(f_2=2f_1\). The system is reciprocal for odd harmonics as $\eta_{ac}(2n+1)=0$, while $\eta$ is reduced for higher even harmonics. The reduction is due to a decrease in the asymmetry of the drive for higher-order harmonics $\eta_{ac}(2n)$, which requires an enhanced driving signal to reach \(\eta=1\) for the chosen parameters. Detailed information about the harmonic dependence of the driving signal asymmetry is provided in the Additional Data section of Supporting Information.
\begin{figure}
    \centering
    \includegraphics[width=1\linewidth]{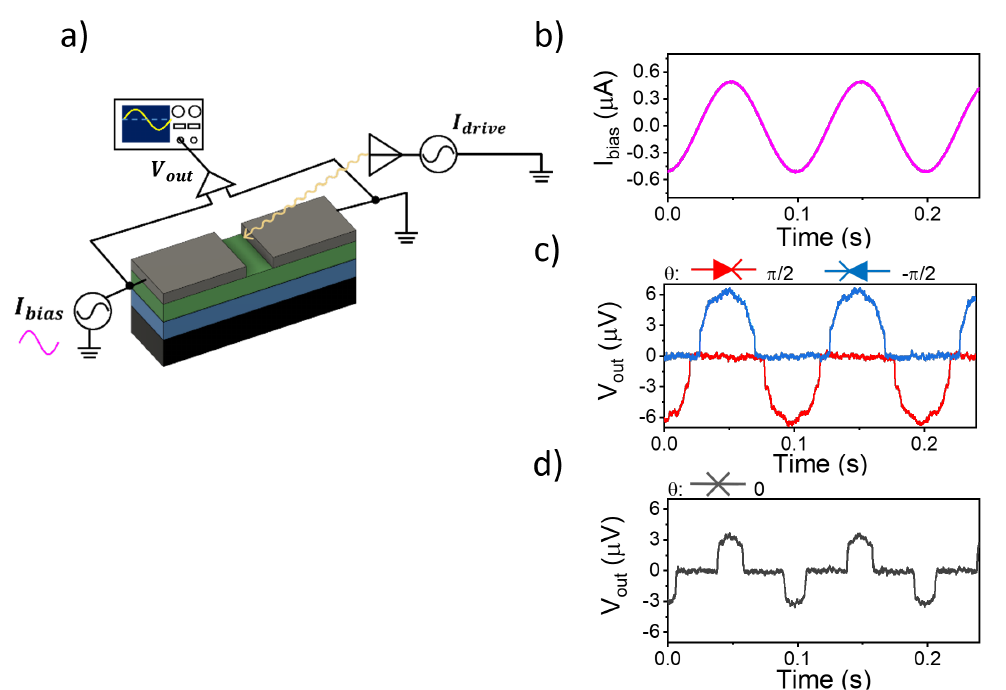}
    \caption{\textbf{Rectification of an AC signal.}  a) Sketch of the experimental setup with driving signal (\(f_1=1.35\) GHz, \(f_2=2f_1\) and emitted power \(P_{RF_1}=-8\) dBm and \(P_{RF_2}=-10\) dBm) through the antenna. The output voltage is real-time measured with an oscilloscope and averaged over 128 waveforms. b) Sinusoidal bias signal of \(f_{AC}=10\) Hz and $I_{bias}^{pp}=1$ \(\mu\)A injected through the JJ. c) Half-wave rectified signals. d) Output signal when no diode effect occurs. Measurements are taken at \(T\approx70\) mK.}
    \label{fig:Figure4}
\end{figure}
\newline
A testbed application of the diode as a rectifier is shown in Figure \ref{fig:Figure4}. 
Here, a low AC bias current ($f_{AC}=10$ Hz $\ll f_{1}$) is injected into the JJ (Figure\ref{fig:Figure4}b) and the output voltage is measured in real time, Figure \ref{fig:Figure4}c-d. 
According to the direction of the diode, tuned by \(\theta\), we observe positive (negative) half-wave rectification of the output voltage as shown in Figure \ref{fig:Figure4}c, for the red (blue) curve. Electrical rectification is not observed at \(\theta=0\) despite to the deformation of the AC output signal due to the non-linear response of the JJ with zero output voltage when the current bias is between \(I_c^-\) and \(I_c^+\) (Figure \ref{fig:Figure4}d).
Moreover, continuous control over \(\theta\) allows intermediate rectification schemes above half-wave rectification.
A full-wave rectifier can be realized by combining multiple superconducting diodes in a bridge structure\cite{ingla-aynes_highly_2024,castellani_superconducting_2025}. Other tested bias frequencies ranging from 30 to 300 Hz are reported in the Supporting Information, Additional Data section.
The current rectification range of this scheme is limited by $I_c$, and in the ideal case ($\eta=1$) this is additionally reduced to $|I_s^*|<I_c$ defined as the critical current in ideal rectification $|I^*_s|=I_c-\text{min}(I_{ac}^+,|I_{ac}^-|)$, as shown in Figure \ref{fig:Figure5}a. Since ideal diode efficiency is achieved when the peak amplitude of the drive matches the critical current, i.e. max$(|I_{ac}^\pm|)=I_c$, variations in $I_c$, which depend on temperature, external magnetic field, and junction length \cite{tinkham_introduction_2004}, will affect the diode efficiency. If the drive is fixed, the efficiency will naturally evolve in response to changes in $I_c$. A discussion on the temperature dependence of $\eta\le1$ is included in Supporting Information, Additional Data section.
\begin{figure}
    \centering
    \includegraphics[width=0.90\linewidth]{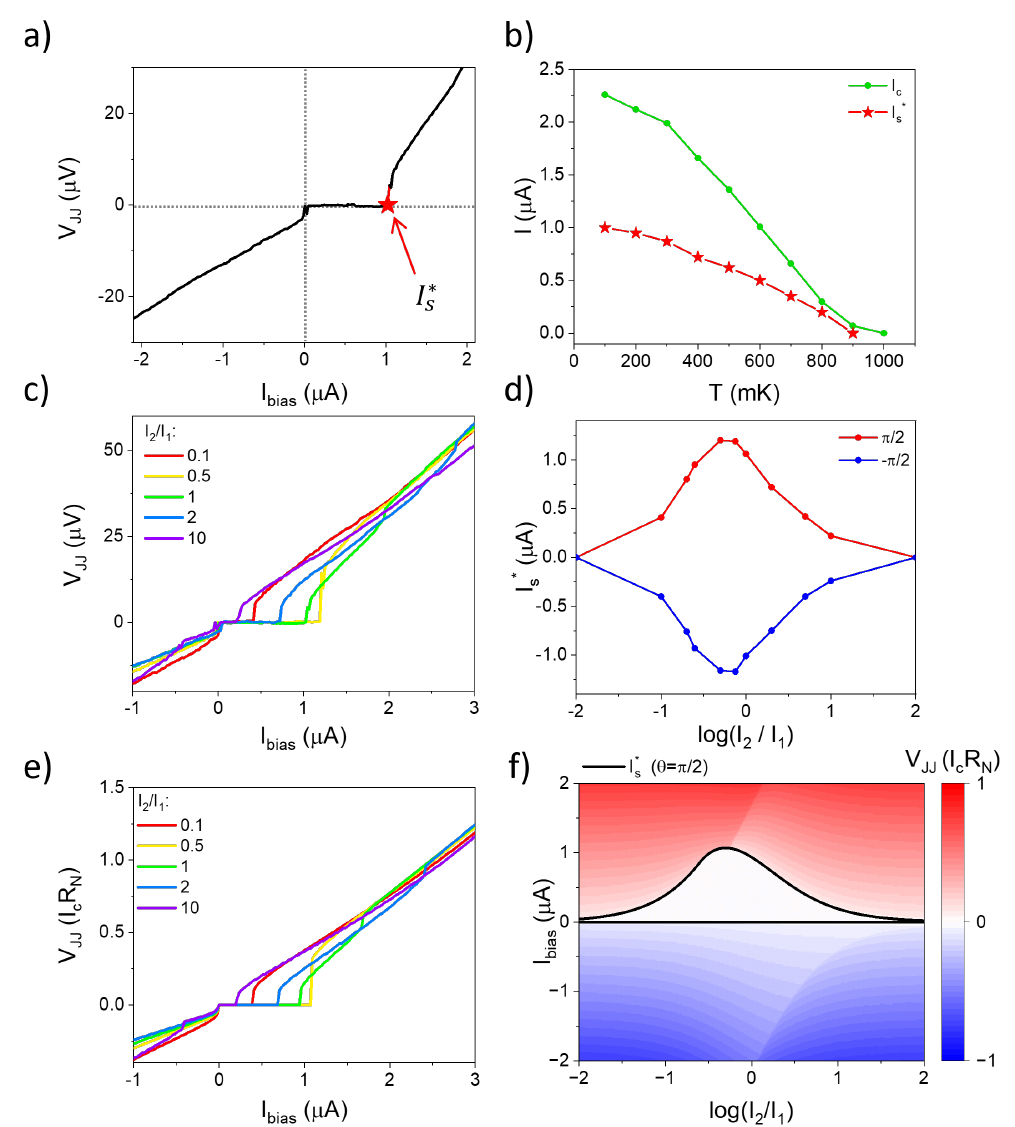}
    \caption{\textbf{Range of operation and limitations: role of temperature and \(I_2/I_1\).} a) VI curve for an ideal diode (driving frequency $f_1=100$ Hz, $I^{pp}=3.8\text{ }\mu\text{A }, I_1=I_2$): \(I_s^*\) is highlighted in red. b) Critical current \(I_c\) (green curve) and \(I_s^*\) (red curve) as a function of the temperature. For the \(I_s^*\)  temperature dependence, a 100 Hz biharmonic drive signal with $I_1=I_2$ at different amplitudes is used. $I_s^*$ is zero for 900 mK. c-d) VI curve (c) and  \(I_s^*\) (d) as a function of \(I_2/I_1\) for \(f_1=100\) Hz. To achieve ideal diode effect $I^{pp}=4.4, 3.6, 3.8, 4.2 \text{ and } 4.6$ $\mu$A respectively for \(I_2/I_1=0.1, 0.5, 1, 2, \text{ and }10\). e-f) Simulated VI curve (e) and colormap (f) as a function of \(I_2/I_1\). For the simulations, we set $R_N=10$ $\hbar/2e^2$ , $C=0.02$ $eI_c/\hbar$ and $2\pi f_1=0.05\omega_c$. The black curve in (f) is the analytical model, where we set $I_{ac}^-=I_c$ and the phase boundary is given by $I_c-I_{ac}^+$. Measurements for (c-d) are taken at $T\approx100$ mK.}
    \label{fig:Figure5}
\end{figure}
\newline
The temperature dependence of \(I_s^*\) naturally follows the damping of \(I_c\), the critical current of the JJ without any applied drive, as reported in Figure \ref{fig:Figure5}b. 
Based on these results, our diode can operate up to temperatures of $\sim$800 mK, where signals up to $I^{pp}=0.5$ \(\mu\)A peak-to-peak can be rectified. We refer to the ability to achieve 100$\%$ diode efficiency even in proximity to the JJ transition temperature as temperature-resistant behavior.
Larger critical currents can be achieved in systems with larger $I_c$, and the limiting temperature depends on the critical temperature of the superconductor and the device geometry.
Therefore, the choice of other superconductors can increase the diode operating conditions.
The impact of the \(I_2/I_1\) ratio on $I_s^*$ is investigated in Figure \ref{fig:Figure5}c , showing the VI characteristics measured at different $I_2/I_1$. The maximum \(I_s^*\) is achieved for \(I_2/I_1=0.5\), as predicted by perturbation theory\cite{scheer_tunable_2025} and suggests a symmetry point around this value as outlined in Figure \ref{fig:Figure5}d where $I_s^*(I_2/I_1)$ is extracted for $\theta=\pm \pi/2$. The diode effect monotonically decreases for larger and smaller ratios.
Figure \ref{fig:Figure5}e shows the simulated VI curves at different \(I_2/I_1\) ratios, confirming the maximum at \(I_2/I_1=0.5\) and the behavior of $I_s^*(I_2/I_1)$, as reported in Figure \ref{fig:Figure5}f (black curve).

It has been demonstrated that a biharmonic drive induces a tunable diode effect within a hybrid Josephson junction, which requires fewer complexities or asymmetries in the system.
Using an antenna enables wireless, fast, and simultaneous control over the diode's direction and efficiency by modifying the driving tones' relative phase. Moreover, the diode exhibits additional tunability by adjusting the relative amplitudes of the two tones.
This diode showcases exceptional adaptability, achieving 100\% efficiency over the drive's broad frequency spectrum, from a few Hz to GHz, and different combinations of even harmonics.
The tunable Josephson diode has effectively rectified AC signals across frequencies and temperatures up to 800 mK, limited only by the superconductor's critical temperature. The rectification range has been tested and optimized for biharmonic drive with a different ratio between the first- and second-harmonic amplitudes.

In summary, we have developed a highly versatile, temperature-resistant, and platform-agnostic superconducting diode that is straightforward to manufacture and can be tuned with easily generated and controlled signals. Notably, the dynamically adjustable polarity of the superconducting diode could facilitate the development of fundamental logic gates and ultra-fast switches, thereby advancing superconductor-based digital electronics. Additionally, the capability of this superconducting diode to function as a controllable active rectifier, essential for the conversion and management of electrical energy, paves the way for advances in superconducting integrated power electronics. Such supercurrent-based rectifiers fundamentally differ from traditional semiconductor rectifiers, leading to reduced power dissipation.
Future investigations will explore the diode in the fast nonadiabatic regime \(2\pi f\sim \omega_c\), an unresolved issue that could extend its operating speed.

\begin{acknowledgement}
The authors thank J. Danon, F. Hassler, D. Scheer and G. De Simoni for valuable help and discussions before and during the development of this work. The authors thank the EU’s Horizon 2020 Research and Innovation Framework Program under Grant Agreement No. 101057977 (SPECTRUM), the PNRR MUR project PE0000023-NQSTI, the Italian project HELICS DFM.AD002.206, the Spanish Comunidad de Madrid (CM) “Talento Program” (Project No. 2022-T1/IND-24070), and the Spanish Ministry of Science, Innovation, and Universities through Grant PID2022-140552NA-I00.

\end{acknowledgement}

\begin{suppinfo}

Further information on device fabrication, experimental setup, and additional measurements studying power, temperature and platform dependencies.

\end{suppinfo}

\newpage
\appendix
\section{Supporting Information}
\section{Methods}
\subsection{Sample information}
The InAsOI heterostructure was grown on GaAs (100) substrate with a Molecular Beam Epitaxy technique. Starting from the 350~\(\mu\)m-thick semi-insulating GaAs (100) substrate, the sample consists of: a 50 nm GaAs buffer, a 100 nm GaAs/AlGaAs superlattice, a 50 nm GaAs layer, a 1.25~\(\mu\)m-thick step-graded \(In_{X}Al_{1-X}As\) metamorphic buffer layer (X from 0.15 to 0.81) that acts as insulator at cryogenic temperature and a 100 nm-thick InAs semiconductive epilayer. This sample structure was adapted from Refs.~\cite{paghi_inas_2025,senesi_structural_2025}.
Classical Hall effect measurements in Hall bar configuration have been used to estimate the sheet carrier concentration and mobility of this sample, respectively: $n_{2D}=1.94\times 10^{12}\mathrm{cm}^{-2}$ and \(\mu _n= 8.5\times 10^3\) cm\(^2\)/Vs at 3K. InAsOI substrates were cleaned (ACE and IPA), passivated with a \((NH_4)_2S_x\) solution and loaded in an electron beam evaporator to evaporate 100 nm of Al all over the sample. In order to define the MESA, an UV-lithography was performed: then, after developing in MF319, the exposed Al layer was etched in Al Etchant Type D and the exposed InAs epilayer was etched in a \(H_3PO_4:H_2O_2\) solution. To define the Josephson junction length, it was necessary to define EBL-markers and deposit a 10/50 nm-thick Ti/Au bilayer in a thermal evaporator; then the Al contacts on the InAs MESA were defined by marker-aligned EBL and selected chemical etching of the Al.
Additional information on device fabrication can be found in \cite{paghi_inas_2025,paghi_josephson_2024,paghi_supercurrent_2024,battisti_extremely_2024}.

\subsection{Low temperature DC and AC setup}

\begin{figure}
    \centering
    \includegraphics[width=0.75\linewidth]{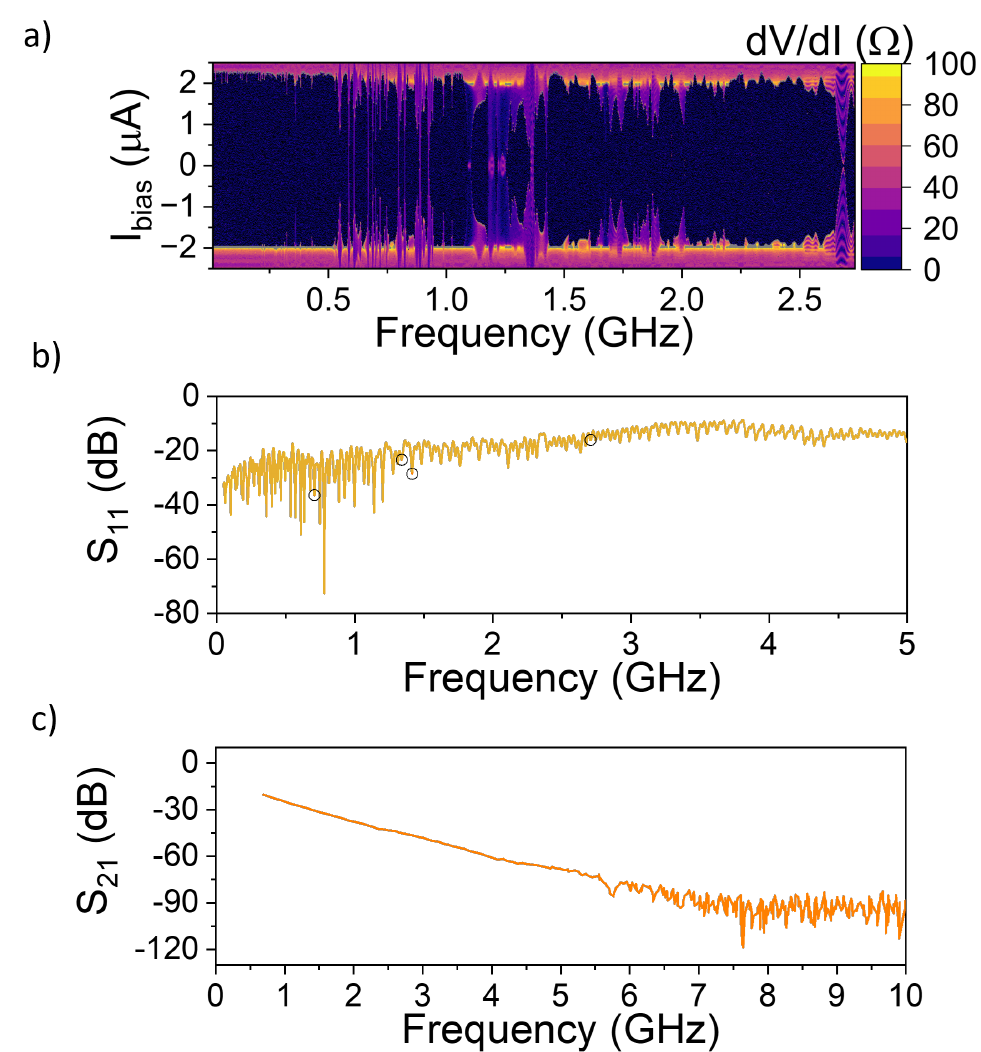}
    \caption{\textbf{Frequency matching of the antenna.} a) Evolution of differential resistance as a function of current bias \(I_{bias}\) and frequency at applied power \(P_{RF}=-10\) dBm. b) Reflection coefficient \(S_{11}\) as a function of the antenna frequency. Circles indicate the chosen frequencies related to the biharmonic-drive diode plots. c) Power received by the antenna relative to the power emitted \(S_{21}\) as a function of the antenna frequency.}
    \label{FigureS1}
\end{figure}

\begin{figure}
    \centering
    \includegraphics[width=0.75\linewidth]{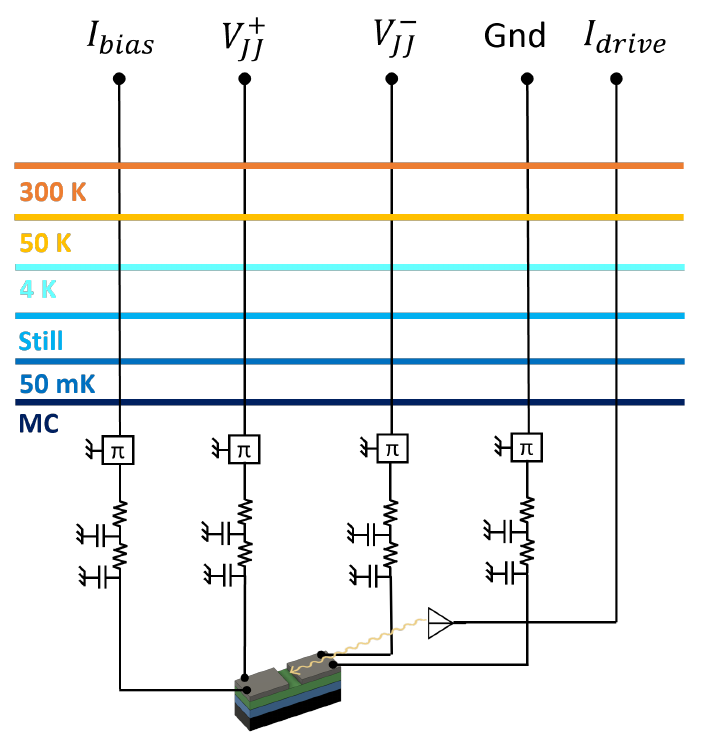}
    \caption{\textbf{DC and microwave setup.} Schematic of the cabling configuration within the dilution refrigerator. RC filters have resistance 1.1 k\(\Omega\)  and capacitance 10 nF: they constitute a low-pass filter block with a cutoff of 10 kHz.}
    \label{FigureS2}
\end{figure}

The device was mounted on a chip in a Leiden cryostat (CF-CS110) equipped with a magnet, cooled down to the base temperature (70 mK). A four-wire configuration was used to carry out the electrical characterization (VI curves) of the device: the junction was current-biased using a Yokogawa GS200 voltage source over a 1 M\(\Omega\) resistor, the voltage \(V_{JJ}\) across the junction was amplified (Voltage Amplifier DL1201) and measured with an Agilent 34401A with NPLC=1. An out-of-plane magnetic field of 60 \(\mu\)T was used only to maximize the switching current and maintained throughout all the measurements: we used a Yokogawa GS200 voltage source over a 100 \(\Omega\) resistor connected to an American Magnetics magnet inside the cryostat.
Microwave drives generated from two Anritsu signal generators (68369B and MG3694A) were summed using a DC pass power splitter/combiner (2-10 GHz, 50 \(\Omega\)) and applied to an open-ended coaxial cable near the device. By fixing the power and varying the frequency, we show that only some emitted frequencies "match" with the response of the device, showing a change in the VI characteristic, as reported in Figure \ref{FigureS1}a: here, the stripes represent a damping of the critical current or the appearance of Shapiro steps. This allowed the selection of the frequencies that optimized the coupling between the antenna and the sample.
Additionally, to characterize the reflection behavior of the antenna we measured with a Vector Network Analyzer the return loss \(S_{11}\), as shown in Figure \ref{FigureS1}b.
The attenuation of the microwave signal due to cables is estimated through the \(S_{21}\) parameter analysis in frequency, as reported in Figure \ref{FigureS1}c: it shows that there is an attenuation of \(-100\) dBm above 5 GHz, so our operational range is below 5 GHz.
For the measurements in Hz and kHz regime, the driving signals were applied with an arbitrary waveform generator (Agilent 33220A) over a 1 M\(\Omega\) resistor, the flowing current was amplified and collected with an Agilent 34401A with NPLC=1.
The real-time measurements were carried out with a similar four-wire technique, using an arbitrary waveform generator (Agilent 33220A) as a source; the voltage drop over the junction and the flowing current were amplified and acquired using a Tektronix TDS 2024B oscilloscope (128 means).
The cabling inside the dilution refrigerator is shown in Figure \ref{FigureS2}: there is a low-pass frequency stage between the Mixing Chamber and the sample.
\newpage
\section{Additional data}
\subsection{Shapiro steps in temperature}
\begin{figure}
    \centering
    \includegraphics[width=1\linewidth]{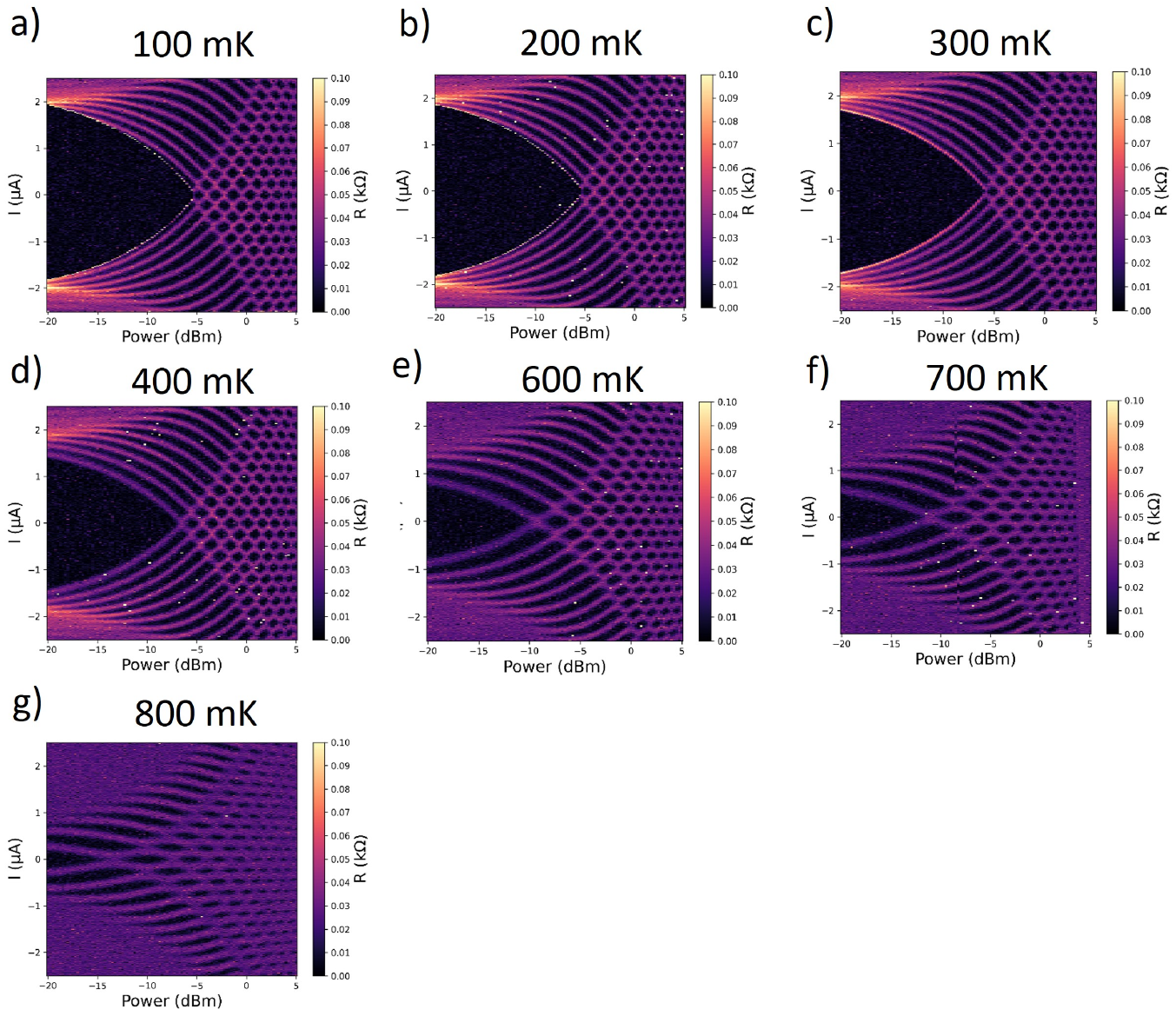}
    \caption{\textbf{Shapiro steps in temperature (\(f=2.7\) GHz).} a-g) Shapiro evolution of differential resistance as a function of current bias (\(I_{bias}\)) and applied power (\(P_{RF}\)) at \(f=2.7\) GHz at different temperatures.}
    \label{FigureS3}
\end{figure}

Shapiro steps were obtained by sweeping the current bias \(I_{bias}\) from negative to positive. Here we display the measurements for a single frequency drive of 2.7 GHz at different temperature (Figure \ref{FigureS3}): Shapiro steps are still observed at 800 mK (Figure \ref{FigureS3}g).
\subsection{Shapiro steps with a biharmonic drive}
\begin{figure}
    \centering
    \includegraphics[width=1\linewidth]{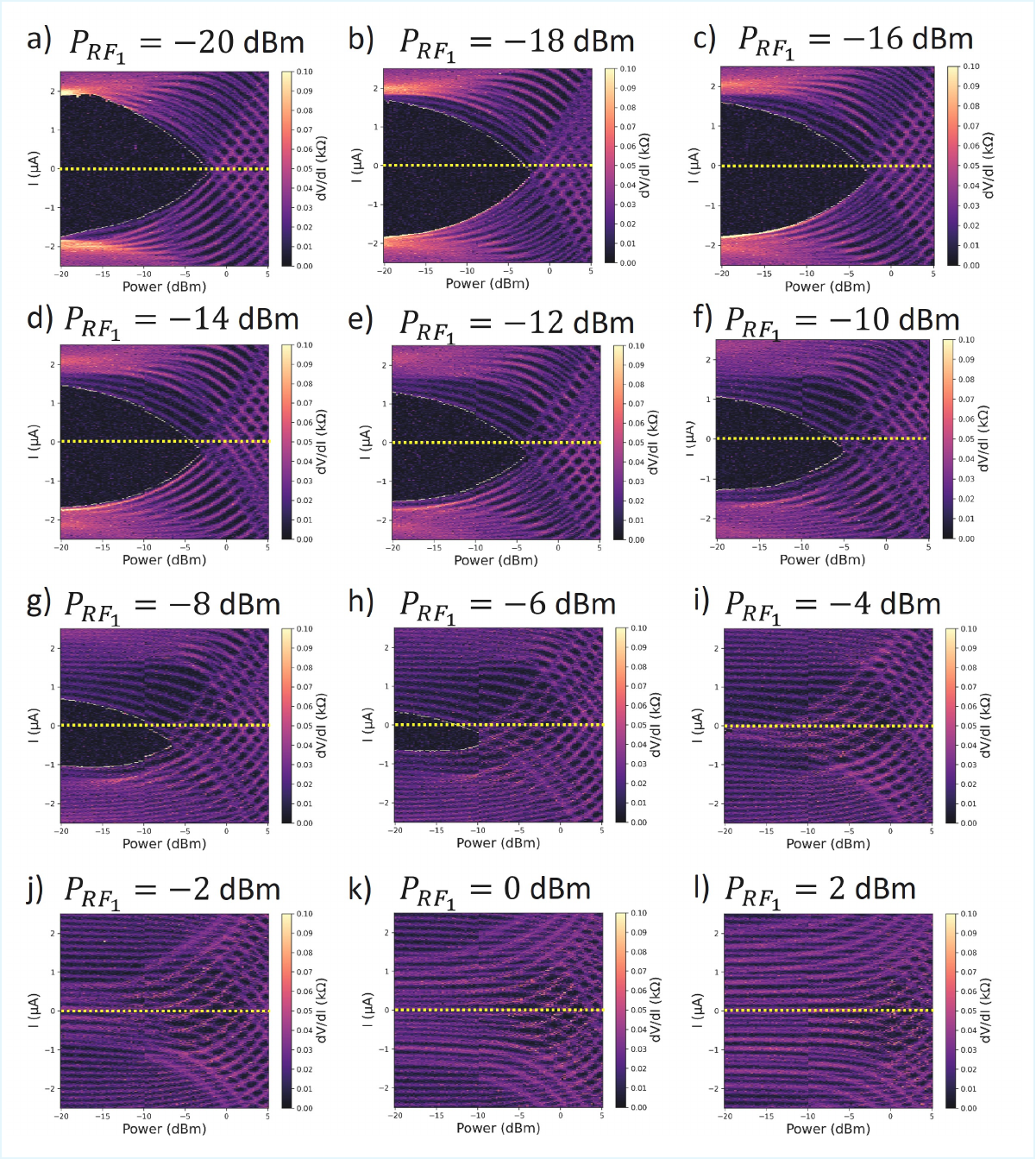}
    \caption{\textbf{Shapiro steps with biharmonic drive.} a-l) Shapiro evolution of differential resistance as a function of current bias (\(I_{bias}\)) and applied power \(P_{RF_2}\) at  \(f_1=1.35\) and \(f_2=2.7\) GHz at different \(I_1\) amplitudes.}
    \label{FigureS4}
\end{figure}

Shapiro steps are also visible when a biharmonic drive (with \(\theta=-\pi /2\))  is applied. In Figure \ref{FigureS4} we show the evolution of the differential resistance as a function of the applied power \(P_{RF_2}\) at different \(P_{RF_1}\).

\subsection{Diode efficiency at different $I_{1,2}$}
\begin{figure}
    \centering
    \includegraphics[width=0.50\linewidth]{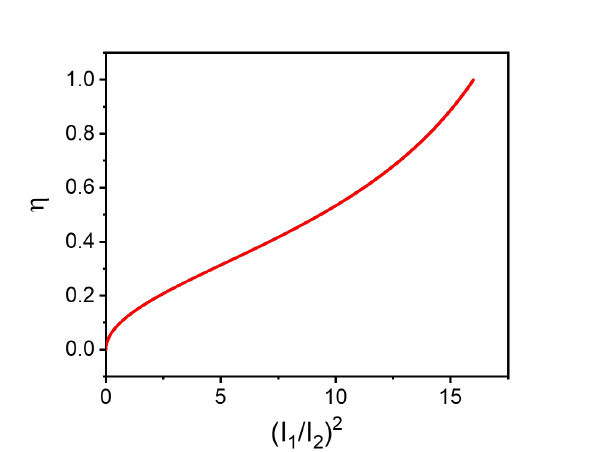}
    \caption{\textbf{Diode efficiency as a function of biharmonic signal amplitudes}. $\eta$ at different $(I_1/I_2)^2$ ratio, for a biharmonic signal with $\theta=\pi/2$, \(f_1=1.35\) GHz and \(f_2=2.7\) GHz. $I_2$ is fixed at $0.2$ $\mu$A.}
    \label{fig:FigureR4}
\end{figure}

\raggedright
Figure \ref{fig:FigureR4} illustrates $\eta$ versus the ratio $(I_1/I_2)^2$ with $I_2$ held constant and $\theta=\pi/2$. The qualitative trend of $\eta$ with respect to $(I_1/I_2)^2$ mirrors the behavior expected as a function of power, assuming a monotonic but nonlinear relationship between power and current and captures the key features of the diode efficiency at different driving amplitudes.

\subsection{Biharmonic drive diode on a different platform (Nb/Au/Nb JJ)}
\begin{figure}
    \centering
    \includegraphics[width=0.90\linewidth]{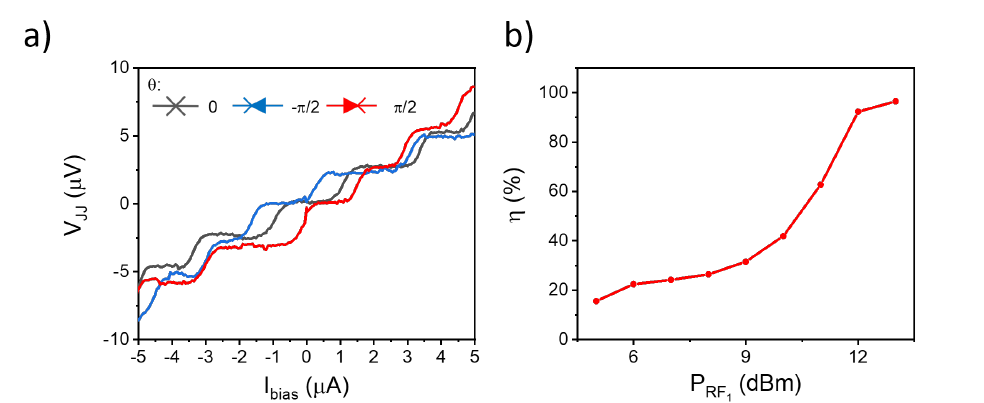}
    \caption{\textbf{Biharmonic drive diode for Nb/Au/Nb JJ.} a) VI curve under applied biharmonic drive with \(f_1=1.35\) GHz and \(f_2=2.7\) GHz at different \(\theta \), \(P_{RF_1}=13\) dBm and \(P_{RF_2}=0\) dBm. b) Diode efficiency at different \(P_{RF_1}\), \(P_{RF_2}=0\) dBm and $\theta=\pi/2$. Measurements are taken at \(T\approx950\) mK.}
    \label{fig:FigureR5}
\end{figure}

\raggedright
The biharmonic-drive diode can be realized on diverse material platforms and does not rely on material-specific spin-orbit interactions or particular device geometry. To further prove the platform-independency of our approach, we performed additional measurements on a Nb/Au/Nb SNS JJ. Figure \ref{fig:FigureR5}a shows the VI curves measured under a biharmonic drive with $\theta=\pm \pi/2,0$, demonstrating the diode effect. The direction and efficiency of the diode are controlled by \(\theta\) ranging from ideal ($\eta=\pm1$ in $\theta=\pm \pi/2$) to zero ($\eta=0$ in $\theta=0$). 
Figure \ref{fig:FigureR5}b shows the evolution of $\eta$ with \(P_{RF_1}\), keeping $P_{RF_2}$ fixed and $\theta=\pi/2$.

\subsection{Harmonic dependence of signal asymmetry}
\begin{figure}
    \centering
    \includegraphics[width=0.50\linewidth]{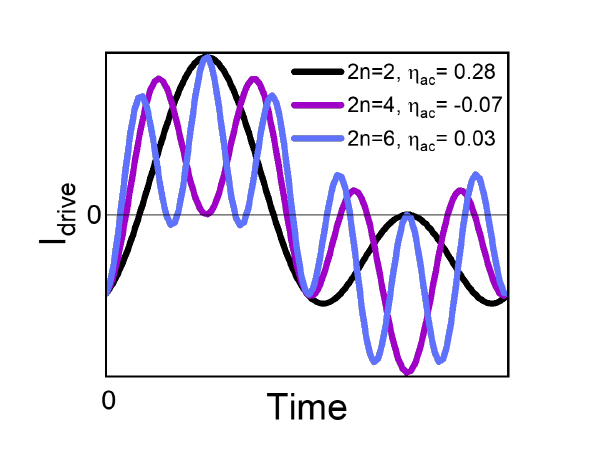}
    \caption{\textbf{Harmonic dependence of signal asymmetry} Biharmonic drive signal $I_{drive}(t)=I_1\sin{(2\pi f_1t)}+I_2\sin{(2\pi (2nf_1)t-\pi/2)}$ and calculated asymmetries $\eta_{ac}$ for $2n=2,4,6$, $I_1=I_2$, $I^{pp}=3.8$ $\mu$A.}
    \label{fig:FigureR1}
\end{figure}

\raggedright
When considering higher order harmonics in the driving signal, the diode efficiency $\eta$ decreases due to the reduction of the drive asymmetry quantified by $\eta_{ac}$, which is directly proportional to $\eta$. As the harmonic order $n$ increases, the waveform extrema become more symmetric, leading to a suppression of $\eta_{ac}$ and a sign reversal. Figure \ref{fig:FigureR1} illustrates this behavior for $2n=2,4,6$, showing the corresponding drive waveforms and calculated $\eta_{ac}$. Increasing the drive amplitudes $I_{1,2}$ can restore the asymmetry for higher harmonics up to the ideal case.

\subsection{Rectification of AC signals}
\begin{figure}
      \centering
      \includegraphics[width=1\linewidth]{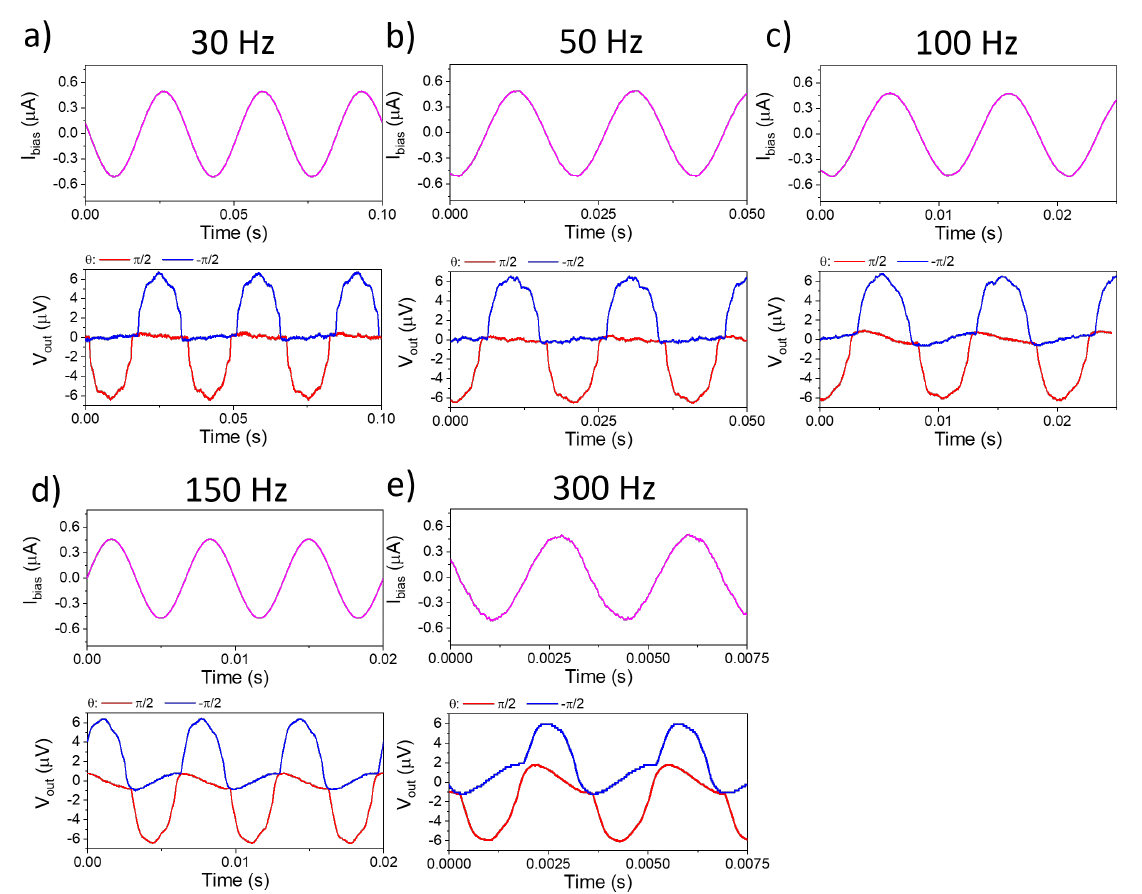}
      \caption{\textbf{Rectification of AC signal (more bias frequencies). } a-e) Sinusoidal bias signal of \(f_{AC}=30-300\) Hz and $I_{bias}^{pp}=1$ \(\mu\)A injected through the JJ and half-wave rectified signal (blue and red curves). The drive signal has\(f_1=1.35\) GHz, \(f_2=2f_1\) and emitted power \(P_{RF_1}=-8\) dBm and \(P_{RF_2}=-10\) dBm}
      \label{fig:FigureS5}
  \end{figure}

The biharmonic-drive diode was used to rectify AC signals: Figure \ref{fig:FigureS5} shows the biharmonic-drive diode as rectifier for bias signals ranging from 30 to 300 Hz. At higher frequency (Figure \ref{fig:FigureS5}e) the output signal is distorted due to the filtering stage in the electrical setup.

\subsection{Role of \(I_2/I_1\) }
\begin{figure}
    \centering
    \includegraphics[width=1\linewidth]{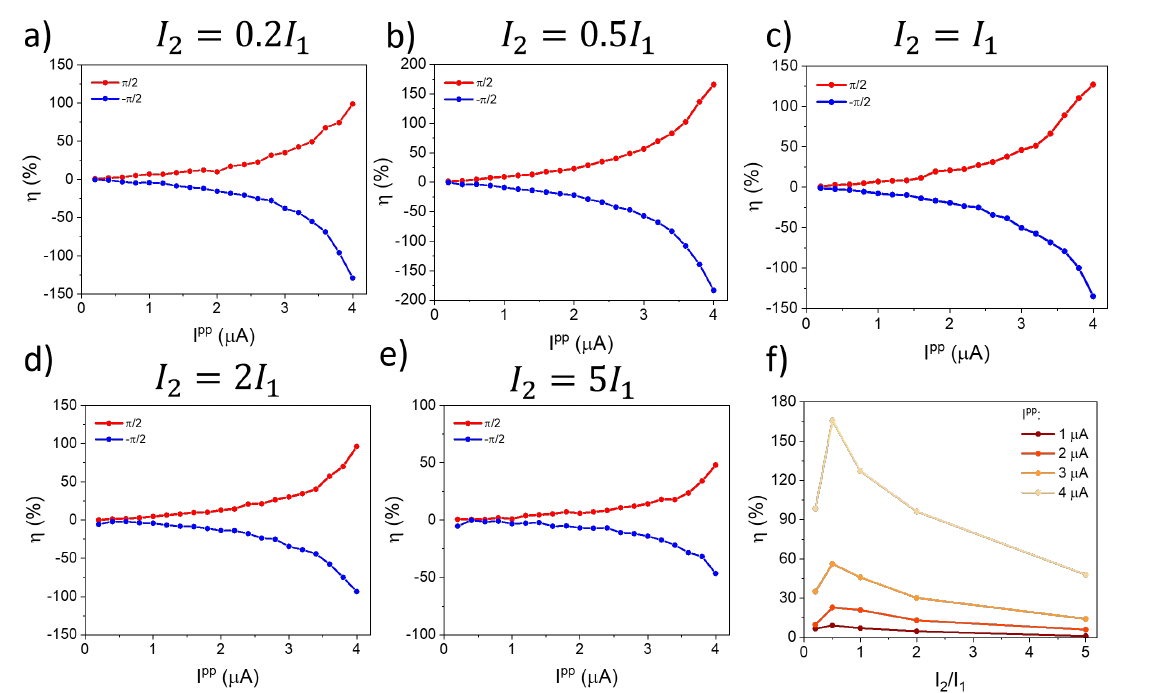}
    \caption{\textbf{Diode efficiency at different \(I_2/I_1\).} a-e) \(\eta\) vs $I^{pp}$ of the driving signal. f) \(\eta\) vs \(I_2/I_1\) at various $I^{pp}$ of the driving signal. Drive frequency \(f_1=100\) Hz.}
    \label{fig:FigureS6}
\end{figure}

Figure \ref{fig:FigureS6} depicts the dependence of the diode efficiency \(\eta\) with respect the ratio  \(I_2/I_1\), that represent the weight of the second harmonic over the first harmonic in the driving signal. The efficiency \(\eta\) is analyzed as a function of peak-to-peak current (\(I^{pp}\)) that we experimentally controlled, once fixed the \(I_2/I_1\) ratio.

\subsection{Temperature dependence of the diode efficiency}
\begin{figure}
    \centering
    \includegraphics[width=0.90\linewidth]{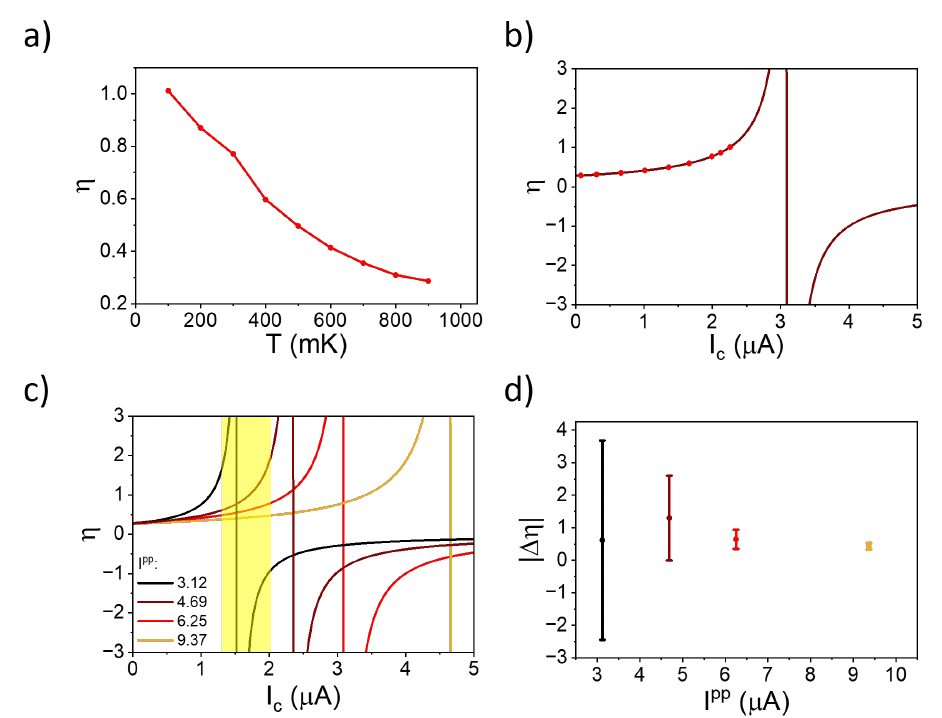}
    \caption{\textbf{Temperature dependence of the diode efficiency} a) $\eta(T)$ calculated with Eq. \eqref{Eq3} and the experimental data of $I_c(T)$. b) $\eta(I_c)$ for a driving signal with $\eta_{ac}=0.27$ and $I^{pp}=6.25$ $\mu$A. The red dots are the value of $\eta$ corresponding at the experimental data of $I_c(T)$. c) $\eta(I_c)$ for driving signals with different $I^{pp}$. The yellow region correspond to a fluctuation of $\Delta I_c\approx0.6$ $\mu$A between 300 and 500 mK. d) Variation of diode efficiency in the yellow region of c) for the different driving signals.}
    \label{fig:FigureR2}
\end{figure}

\begin{figure}
    \centering
    \includegraphics[width=0.90\linewidth]{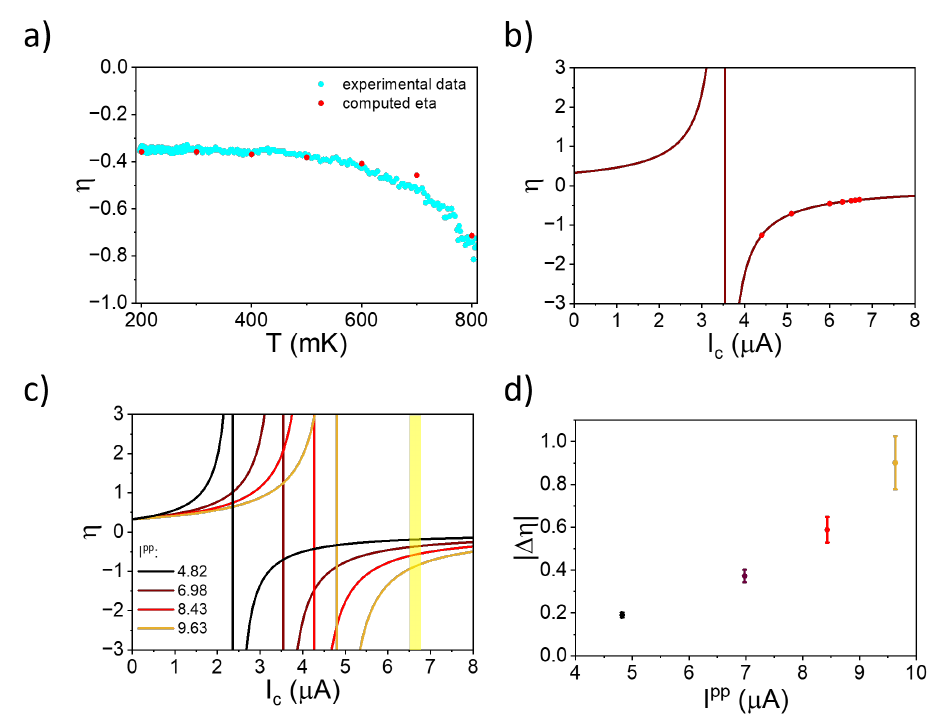}
    \caption{\textbf{Temperature dependence of the diode efficiency for Nb/Au/Nb JJ.} a) $\eta(T)$ measured (cyan) and calculated (red) with Eq. \eqref{Eq3} and the experimental data of $I_c(T)$. b) $\eta(I_c)$ for a driving signal with $\eta_{ac}=0.33$ and $I^{pp}=8.43$ $\mu$A. The red dots are the value of $\eta$ corresponding at the experimental data of $I_c(T)$. c) $\eta(I_c)$ for driving signals with different $I^{pp}$. The yellow region correspond to a fluctuation of $\Delta I_c\approx0.2$ $\mu$A between 300 and 500 mK. d) Variation of diode efficiency in the yellow region of c) for the different driving signals.}
    \label{fig:FigureR3}
\end{figure}

\raggedright
According to
\begin{equation}
\eta=\frac{\eta_{ac}}{1-\frac{2I_c}{I_{ac}^++|I_{ac}^-|}},
\label{Eq3}
\end{equation}
the diode efficiency $\eta$ naturally varies with temperature, because $I_c(T)$. Figure \ref{fig:FigureR2}a and Figure \ref{fig:FigureR2}b respectively illustrate $\eta(T)$ and $\eta(I_c)$, using a biharmonic drive calibrated to yield a diode efficiency of $100\%$ at 100 mK and the experimental data of $I_c(T)$. For a fixed temperature (and therefore for a fixed $I_c$), the diode efficiency can be tuned by changing the driving parameters as illustrated in Figure \ref{fig:FigureR2}, which shows $\eta(I_c)$ for different $I^{pp}=I_{ac}^++|I_{ac}^-|$ of the driving signal. For a fluctuation $\Delta T=100$ mK around 400 mK the critical current of the device changes approximately of $\Delta I_c\approx0.6$ $\mu$A. As shown in Figure \ref{fig:FigureR2}d, different choices of the driving parameters will result in different values of $\Delta \eta$, showing that an appropriate selection of the driving parameters can effectively mitigate the sensitivity of $\eta$ to fluctuations in temperature and $I_c$ enhancing operational stability and robustness of the diode.
Additionally we conducted another experiment with a different device consisting in a Nb/Au/Nb JJ, for which we investigated intermediate configurations of the diode effect. In Figure \ref{fig:FigureR3}a and Figure \ref{fig:FigureR3}b, we show experimental and computed data for $\eta(T)$ and $\eta(I_c)$, respectively. This time, we consider an intermediate diode configuration, so that at 200 mK the diode efficiency is $35\%$. Also for this device we show how $\eta$ evolves with the driving parameters (Figure \ref{fig:FigureR3}c) and with fluctuations of $I_c$ (Figure \ref{fig:FigureR3}d).”

\newpage
\providecommand{\latin}[1]{#1}
\makeatletter
\providecommand{\doi}
  {\begingroup\let\do\@makeother\dospecials
  \catcode`\{=1 \catcode`\}=2 \doi@aux}
\providecommand{\doi@aux}[1]{\endgroup\texttt{#1}}
\makeatother
\providecommand*\mcitethebibliography{\thebibliography}
\csname @ifundefined\endcsname{endmcitethebibliography}  {\let\endmcitethebibliography\endthebibliography}{}

\end{document}